\documentstyle[12pt]{article}

\def\gt{{\tilde{\gamma}}}

\def\GeV{\,{\rm GeV}}

\def\MeV{\,{\rm MeV}}
\def\sec{\,{\rm sec}}

\def\cmm2{{\,\rm cm^{-2}}}
\def\cm2{{\,{\rm cm}^2}}
\def\cmm3{{\,{\rm cm}^{-3}}}
\def\gcmm3{{\,{\rm g\,cm^{-3}}}}

\def\mpl{{m_{\rm Pl}}}

\def\la{\mathrel{\mathpalette\fun <}}
\def\ga{\mathrel{\mathpalette\fun >}}
\def\fun#1#2{\lower3.6pt\vbox{\baselineskip0pt\lineskip.9pt
  \ialign{$\mathsurround=0pt#1\hfil##\hfil$\crcr#2\crcr\sim\crcr}}}

\begin{document}
\baselineskip=16pt
\pagestyle{empty}
\begin{center}
\rightline{FERMILAB--Pub--93/181-A}
\rightline{astro-ph/9307***}
\rightline{submitted to {\it Physical Review D}}
\vspace{.2in}

{\bf EFFECT OF FINITE NUCLEON MASS ON PRIMORDIAL NUCLEOSYNTHESIS}\\
\vspace{.2in}

Geza Gyuk$^{1}$ and Michael S. Turner$^{1,2,3}$ \\
\vspace{.1in}

$^1${\it Department of Physics,\\
The University of Chicago, Chicago, IL  60637-1433}\\
$^2${\it NASA/Fermilab Astrophysics Center,
Fermi National Accelerator Laboratory, Batavia, IL  60510-0500}\\
$^3${\it  Department of Astronomy \& Astrophysics,\\
Enrico Fermi Institute,
The University of Chicago, Chicago, IL 60637-1433}\\
\end{center}

\vspace{.3in}

\centerline{\bf ABSTRACT}
\bigskip

\noindent We have modified the standard code for
primordial nucleosynthesis to include the
effect of finite nucleon mass on the weak-interaction
rates as calculated by Seckel \cite{seckel}.
We find a small, systematic increase in the $^4$He yield,
$\Delta Y \simeq 0.0057\,Y$, which is insensitive
to the value of the baryon-to-photon ratio $\eta$
and slightly larger than Seckel's estimate.
The fractional changes in the abundances of D, $^3$He,
and $^7$Li range from 0.08\% to 3\% for
$10^{-11}\le \eta \le 10^{-8}$.

\newpage
\pagestyle{plain}
\setcounter{page}{1}
\section{Introduction}

Primordial nucleosynthesis is one of the cornerstones
of the hot big-bang cosmology.  The agreement between
its predictions for the abundances of D, $^3$He, $^4$He
and $^7$Li and their inferred primordial abundances provides its
earliest, and perhaps most, stringent test.
Further, big-bang nucleosynthesis has been used to provide
the best determination of the baryon density \cite{ytsso,walker}
and to test particle-physics theories, e.g., the stringent limit
to the number of light neutrino species \cite{nulimit}.

The scrutiny of primordial nucleosynthesis, both on
the theoretical side and on the observational side, has been constant:
Reaction rates have been updated and the effect
of their uncertainties quantified \cite{rates},
finite-temperature corrections have been taken into
account \cite{dicusetal}, the effect of inhomogeneities in the
baryon density explored \cite{matthews}, and the
slight effect of the heating of neutrinos by $e^\pm$ annihilations
has been computed \cite{dt}; the primordial abundance
of $^7$Li has been put on a firm basis \cite{lithium}, the production
and destruction of D and $^3$He have been studied carefully
\cite{dhe3}, and astrophysicists now argue about the third significant
figure in the primordial $^4$He abundance \cite{he4}.

A measure of the progress in this endeavour is
provided by the shrinking of the ``concordance region''
of parameter space.  The predicted and measured primordial
abundances agree provided:  the baryon-to-photon ratio
lies in the narrow interval $3 \times 10^{-10}\la \eta \la
4\times 10^{-10}$ and the equivalent number of light
neutrino species $N_\nu \la 3.3$ \cite{walker}.
The shrinking of the concordance interval
motivates the study of smaller and smaller effects.

Seckel \cite{seckel} has recently calculated the corrections to
the weak-interaction rates that arise from taking
account of the finite nucleon mass (in the standard code
these rates are computed in the infinite-nucleon-mass
limit).  The corrections involve terms of
order $m_e/m_N$, $T/m_N$, and $Q/m_N$, which are all of the
order of 0.1\%.  Here $m_e$ is the electron mass, $m_N$ is
the nucleon mass, $T\sim {\cal O}(\MeV )$ is the temperature during
the epoch of nucleosynthesis, and $Q=m_n-m_p=1.293\MeV$ is the
neutron-proton mass difference.  The weak interaction
rates govern the neutron-to-proton ratio and thereby
are crucial to the outcome of nucleosynthesis; e.g.,
the mass fraction of $^4$He produced is roughly twice
the neutron fraction at the time nucleosynthesis commences
($T\sim 0.07\MeV$).

The net effect of the finite-nucleon-mass corrections
is to decrease the weak rates by about 1\% around the time
of nucleosynthesis.  Based upon a simple
code that follows the neutron fraction Seckel estimated
that the cumulative effect of all the corrections increase
the mass fraction of $^4$He synthesized by $\Delta Y \approx
0.0012$.  Because the third significant figure of the
primordial $^4$He abundance is now very relevant, we
decided to incorporate the finite-nucleon-mass corrections to the
weak-interaction rates into the standard nucleosynthesis code
\cite{kawano}.  In the next Section we describe the modifications we made;
we finish with a discussion of our results for
the change in the yield of $^4$He, which is slightly larger than
Seckel's estimates, and for the changes in the
yields of the other light elements.

\section{Modifications to the Standard Code}

\subsection{Role of weak interactions}

The weak interactions that interconvert neutrons and
protons, $n\leftrightarrow p + e +\nu$, $n+e\leftrightarrow
p+\nu$, and $n+\nu \leftrightarrow p+e$, play a crucial
role as they govern the neutron fraction, and the neutron fraction
ultimately determines the amount of
nucleosynthesis that takes place.  (Here
and throughout we use $e$ to indicate electron or positron,
and $\nu$ to indicate electron neutrino or antineutrino;
the appropriate particle or antiparticle designation follows from
charge and lepton number conservation.)

The weak-interaction rate per nucleon is very roughly
$\Gamma_{n\rightarrow p}\sim \Gamma_{p\rightarrow n}
\sim G_F^2T^5$, while the expansion rate of the Universe $H\sim
T^2/\mpl$; here $G_F = 1.1664\times 10^{-5}\GeV^{-2}$
is the Fermi constant and $\mpl =1.22\times 10^{19}\GeV$
is the Planck mass.  At temperatures greater than about $0.8\MeV$,
$\Gamma$ is greater than $H$, and the neutron-to-proton
ratio tracks its equilibrium value
\begin{equation}
\left({n\over p}\right)_{\rm EQ} =
(m_n/m_p)^{3/2}e^{-Q/T}.
\end{equation}
Two comments are in order. First, in the infinite-nucleon-mass
limit the prefactor is unity; taking account of this factor tends to increase
the equilibrium ratio by about 0.2\%, suggesting
that the neutron abundance and final $^4$He abundance
should be correspondingly larger.  Second, at a temperature
of about $1\MeV$, the neutrino and photon temperatures begin
to deviate as neutrinos decouple from the electromagnetic
plasma ($e^\pm$, $\gamma$) and $e^\pm$ annihilations begin
to heat the photons relative to the neutrinos as $e^\pm$ pairs
transfer their entropy to the photons.  When this happens
the equilibrium value of the neutron-to-proton ratio is no
longer given by such a simple formula.

When the temperature of the Universe drops below about
$0.8\MeV$ the weak-interaction rate are no longer greater
than the expansion rate and the neutron-to-proton ratio ceases
to track its equilibrium value, and is said to ``freeze out.''
Until nucleosynthesis begins in earnest ($T\sim 0.07\MeV$)
the neutron-to-proton ratio decreases slowly due to weak
interactions (especially neutron decay).  For $\eta\sim
3\times 10^{-10}$ it decreases from about 1/6 at freeze
out to about 1/7 when nucleosynthesis begins, finally resulting
in a mass fraction of $^4$He equal to about 25\%.  For
a detailed discussion of the physics of primordial nucleosynthesis
see Refs.~\cite{Weinberg,KT}.

In the infinite-nucleon-mass limit the rates (per particle) for
the six reactions that interconvert neutrons and protons
are given by \cite{Weinberg}
\begin{eqnarray}\label{eq:weak}
\lambda (n+\nu \rightarrow p+e) & = &{\cal C}\,\int_Q^\infty\,
{p\,E\,(E-Q)^2\,dE \over
[e^{(E-Q)/T_\nu} + 1]\,[e^{-E/T} +1]}  ;\nonumber\\
\lambda (n+ e \rightarrow p+\nu ) & = & {\cal C}\,\int_{m_e}^\infty\,
{p\,E\,(E+Q)^2\,dE \over
[e^{-(E+Q)/T_\nu} + 1]\,[e^{E/T} +1]}  ;\nonumber\\
\lambda (n \rightarrow p+e+\nu ) & = & {\cal C}\,\int_{m_e}^Q\,
{p\,E\,(E-Q)^2\,dE \over
[e^{(E-Q)/T_\nu} + 1]\,[e^{-E/T} +1]}  ;\nonumber\\
\lambda (p+e\rightarrow n+\nu ) & = & {\cal C}\,\int_Q^\infty\,
{p\,E\,(E-Q)^2\,dE \over
[e^{(Q-E)/T_\nu} + 1]\,[e^{E/T} +1]}  ;\nonumber\\
\lambda (p+\nu \rightarrow n+e) & = & {\cal C}\,\int_{m_e}^\infty\,
{p\,E\,(E+Q)^2\,dE \over
[e^{(E+Q)/T_\nu} + 1]\,[e^{-E/T} +1]}  ;\nonumber\\
\lambda (p+e+\nu \rightarrow n) & = & {\cal C}\,\int_{m_e}^Q\,
{p\,E\,(E-Q)^2\,dE \over
[e^{(Q-E)/T_\nu} + 1]\,[e^{E/T} +1]}  ;
\end{eqnarray}
where $T$ denotes the photon temperature,
$T_\nu$ the neutrino temperature, and the common factor
$${\cal C} = {G_F^2 \cos^2\theta_C (1+3c_A^2)\over
2\pi^3}\,f_{\rm EM}.$$
Here $\theta_C$ is
the Cabibbo angle ($\cos\theta_C =0.975$), $c_A = 1.257$
is the ratio of the axial vector to vector coupling of the
nucleon, and $f_{\rm EM} \sim 1.08$ quantifies the radiative
and Coulomb corrections, which for our purposes here
are not important (for more details see Ref.~\cite{dicusetal}).

\subsection{Finite-mass corrections}

Seckel has calculated the corrections to the weak rates
due to finite nucleon mass \cite{seckel}.  He has grouped them into
three categories; the corrections due to:  (i) thermal motion
of the target nucleon; (ii) the recoil energy
of the outgoing nucleon; and (iii) weak magnetism.
The net effect of these corrections is to reduce the
weak rates by about 1\% around the time the neutron-to-proton
ratio freezes out.  This in turn causes freeze out to occur
slightly earlier, at a higher value of the neutron-to-proton
ratio, resulting in an increase in $^4$He production.

The corrections to the rates for neutron decay and inverse
decay are the simplest, just the factor that accounts for time
dilation,
\begin{equation}
\lambda (n\leftrightarrow p+e+\nu ) \longrightarrow
\left( 1 -{3\over 2}{T\over m_N}\right) \lambda (n\leftrightarrow p+e +\nu ).
\end{equation}
Since the neutron lifetime is used to normalize all the
weak rates, the recoil effect is automatically taken into account.

The corrections to the weak interactions that involve
$2\leftrightarrow 2$ scatterings are organized
in a way such that the integrands in the previous expressions,
cf. Eqs.~(\ref{eq:weak}), are simply multiplied
by a correction factor,\footnote{Our notation differs very
slightly from that of Ref.~\cite{seckel}; $\gt_i = \gamma_i/\gamma_0$.}
\begin{equation}
1-\delta_n + \gt_{\rm wm} + \gt_{\rm rec} + \gt_{\rm th}
+\gt_{\rm etc},
\end{equation}
where $\delta_n = -0.00201$ ``uncorrects'' the neutron lifetime
for recoil effects, which are taken into account in the
term $\gt_{\rm rec}$, $\gt_{\rm wm}$ is the weak magnetism
correction, $\gt_{\rm th}$ is the thermal correction, and
$\gt_{\rm etc}$ is the sum of three smaller (by a factor
of 100) and less important terms, which we have included
in our computations, but which have negligible effect.

The weak magnetism, recoil, and thermal corrections are given by \cite{seckel}
\begin{eqnarray}
\gt_{\rm wm} & = & {2c_Af_2\over 1+3c_A^2}\,
{E_3k_1^2+E_1k_3^2\over m_NE_1E_3}; \\
\gt_{\rm rec} & = & {\gt_{\rm wm}\over f_2}
 -{2E_1k_3^2 +E_3(k_1^2+k_3^2) \over 2m_Nk_3^2}
 {m_1^2-m_3^2-Q^2 \over 2(1+3c_A^2)m_NE_3}  \nonumber\\
 & \  & +{c_A^2\over 1+3c_A^2}\,{6E_1^2E_3 -6E_1E_3^2
 -3E_1k_1^2-4E_3k_1^2-E_1k_3^2 \over 2m_NE_1E_3} ;\\
\gt_{\rm th} & = & \left({T\over m_N}\right)\,
\left[ {3E_1^2 +2k_1^2\over 2E_1E_3} +{3k_1^2E_1 +3E_1^2E_3
+2k_1^2E_3 \over 2E_1k_3^2} -{k_1^2E_3^2\over 2k_3^4}\right];
\end{eqnarray}
subscript 1 (3) refers to the incoming (outgoing) lepton,
subscript 2 (4) refers to the incoming (outgoing) nucleon,
$E_i$ is the energy of the $i$th particle, $k_i$ is its
momentum, and $f_2=\pm 3.62$ where $+$ applies to reactions
with leptons and $-$ to reactions with antileptons.
For the three smaller terms embodied in $\gt_{\rm etc}$
we refer the reader to Ref.~\cite{seckel}.

Three small points; the limits of integration are those
in the infinite-nucleon-mass limit, cf.~Eqs.~(\ref{eq:weak}),
with one exception (see below).  In Eqs. (\ref{eq:weak})
the integrals are performed over the energy of the electron; the
neutrino energy is just that of the electron plus or minus
$Q$.  The integral of the final term in $\gt_{\rm th}$,
which is proportional
to $k_3^{-4}$, diverges for the reaction $p+\nu \rightarrow
n+e$ for the infinite-nucleon-mass limits because
$k_3\rightarrow 0$ at threshold.  For this
term, and only this term, the lower limit of integration
involves the finite-nucleon-mass kinematics:  the minimum
electron momentum is:  $(k_3)_{\rm min} = (m_e/m_N)^{1/2}(Q+m_e)$
\cite{seckel2}.  Because essentially all of the integral
accumulates near $(k_3)_{\rm min}$ this term can be integrated
analytically:
\begin{equation}
\left({T\over m_N}\right)\int {p\,E\,(E+Q)^2\,dE\over [e^{(E+Q)/T_\nu}+1]
[e^{-E/T}+1]} {k_1^2E_3^2\over 2k_3^4}
= {(T/m_N)(Q+m_e)^3m_e^{3/2}m_N^{1/2}\over 2
[e^{(Q+m_e)/T_\nu}+1][e^{-m_e/T}+1]} .
\end{equation}

Finally, we mention that Seckel \cite{seckel} has derived linear fits
to the perturbed weak-interaction scattering rates:
\begin{eqnarray}
\lambda (n\rightarrow p) & \rightarrow & \left[ 1 -\delta_n -0.00185
-0.01032\left({T\over m_N}\right)\right]\lambda (n\rightarrow p);\\
\lambda (p\rightarrow n) & \rightarrow & \left[ 1-\delta_n +0.00136
-0.01067\left({T\over m_N}\right)\right]\lambda (p\rightarrow n);
\end{eqnarray}
where the first linear correction applies to the
$2\leftrightarrow 2$ reactions that convert neutrons to protons,
$n+\nu\rightarrow p+e$ and $n+e\rightarrow p+\nu$, and
is valid for $2\MeV \ga T \ga 0.3 \MeV$, and the
second applies to the $2\leftrightarrow 2$ reactions that convert protons
to neutrons, $p+\nu\rightarrow n+e$
and $p+e\rightarrow n+\nu$, and is valid for $2\MeV\ga T \ga 0.7\MeV$.

\section{Results and Conclusions}

We have modified the integrands in the subrountines that compute
the weak-interaction rates in the standard nucleosynthesis
code \cite{kawano} as outlined above.  For comparison,
we have also modified the standard code just using Seckel's
linear fit, which is much easier to implement since the
usual rate is multiplied by a factor outside the integral.
Our results, which were obtained by taking three massless
neutrino species and a mean neutron lifetime of $889\sec$, are
shown in Figs.~1 and 2.

In Fig.~1 the change in the mass fraction of $^4$He and
0.0057 times the $^4$He yield are shown.  While
over the range $\eta = 10^{-11}$ to $10^{-8}$ $\Delta Y$
varies from about 0.0004 to 0.0015, $\Delta Y$ is remarkably
close to $0.0057\,Y$.
We also computed the change in $^4$He yield using Seckel's linear fit;
for the values of $\eta$ above, $\Delta Y/Y$ was slightly
lower, by about 0.05\%.
Neglecting the subdominant terms ($\gt_{\rm etc}$) decreases
$\Delta Y/Y$ from about 0.57\% to about 0.53\%.

In Fig.~2 we show the fractional changes in the abundances
of D, $^3$He, and $^7$Li (relative to H).  These
changes, over the same interval in baryon-to-photon
ratio, range from 0.08\% to almost 3\%.  Since the inferred
primeval abundances of these elements are no where near as
well known as that of $^4$He, these changes
are of little relevance at present.

\vskip 1.5cm
\noindent  We thank David Seckel
for many helpful conversations.  This work was supported in part by the
DOE (at Chicago and Fermilab), by the NASA through
NAGW-2381 (at Fermilab), and GG's NSF predoctoral fellowship.

\newpage

\begin{center}
{\bf FIGURE CAPTIONS}
\end{center}
\medskip

\noindent {\bf Figure 1:}  The change in the yield
of $^4$He as a function of the baryon-to-photon ratio $\eta$,
and, for comparison, 0.0057 times the $^4$He mass fraction.

\bigskip
\noindent{\bf Figure 2:}  The fractional changes in the yields of
D (solid curve), $^3$He (broken curve), and $^7$Li (dotted curve)
as a function of the baryon-to-photon ratio $\eta$.

\end{document}